\documentclass[%
 reprint,
 superscriptaddress,
 amsmath,amssymb,
 aps,
 prl,
]{revtex4-1}

\usepackage{graphicx}% Include figure files
\usepackage{dcolumn}% Align table columns on decimal point
\usepackage{bm}% bold math
\usepackage{epsfig,graphics,amsmath,amssymb}
\usepackage{epstopdf}
\usepackage{upgreek}
\usepackage{booktabs}
\usepackage{color}
\usepackage{pifont}
\usepackage{float}

\usepackage{hyperref}% add hypertext capabilities
\hypersetup{
	hypertex=true,
	colorlinks=true,
	linkcolor=blue,
	filecolor=blue,
	urlcolor=blue,
	citecolor=blue,
}
\begin{document}

\preprint{APS/123-QED}

\title{Catching the geometric phase effect around conical intersection in molecules by high order harmonic spectroscopy}

\author{Guanglu Yuan}
\affiliation{Institute of Ultrafast Optical Physics, MIIT Key Laboratory of Semiconductor Microstructure and Quantum Sensing, Department of Applied Physics, Nanjing University of Science and Technology, Nanjing 210094, China}
\affiliation{State Key Laboratory of Precision Spectroscopy, East China Normal University, Shanghai 200062, China}

\author{Ruifeng Lu}
\thanks{rflu@njust.edu.cn}
\affiliation{Institute of Ultrafast Optical Physics, MIIT Key Laboratory of Semiconductor Microstructure and Quantum Sensing, Department of Applied Physics, Nanjing University of Science and Technology, Nanjing 210094, China}

\author{Shicheng Jiang}
\thanks{scjiang@lps.ecnu.edu.cn}
\affiliation{State Key Laboratory of Precision Spectroscopy, East China Normal University, Shanghai 200062, China}

\author{Konstantin Dorfman}
\thanks{dorfmank@lps.ecnu.edu.cn}
\affiliation{State Key Laboratory of Precision Spectroscopy, East China Normal University, Shanghai 200062, China}
\affiliation{Collaborative Innovation Center of Extreme Optics, Shanxi University, Taiyuan, Shanxi 030006, China}
\affiliation{Himalayan Institute for Advanced Study, Unit of Gopinath Seva Foundation, MIG 38, Avas Vikas, Rishikesh, Uttarakhand 249201, India}

\date{\today}% It is always \today, today,

\begin{abstract}
Nonadiabatic dynamics around an avoid crossing or a conical intersection play a crucial role in the photoinduced processes of most polyatomic molecules. The present work shows that the topological phase in conical intersection makes the behavior of pump-probe high-order harmonic spectroscopy different from the case of avoid crossing. The coherence built up when the system crosses the avoid crossing will lead to the oscillatory behavior of the spectrum, while the geometric phase erodes these oscillations in the case of conical intersection. Additionally, the dynamical blueshift and the splitting of time-resolved spectrum allow capturing the snapshot dynamics with sub-femtosecond resolution.
\end{abstract}

\maketitle
%\tableofcontents
In polyatomic molecules, the avoid crossing (AC) and conical intersection (CI) play important roles in various photophysical and photochemical dynamical processes \cite{domcke2012role,worth2004beyond}. Benefiting from the progress in the laser technology, many optical methods have been developed to probe the nonadiabatic dynamics in the vicinity of CI \cite{zinchenko2021sub,hosseinizadeh2021few,nam2021conical,timmers2019disentangling,neville2018ultrafast,kowalewski2017simulating}. Attosecond transient-absorption spectroscopy succeeded in mapping the AC dynamics directly \cite{kobayashi2019direct}, and later achieved sub-7-femtosecond resolution in CI by extending the attosecond pulse to carbon K-edge \cite{zinchenko2021sub}. A background-free technique called TRUECARS (Transient redistribution of ultrafast electronic coherences in attosecond Raman signals) \cite{kowalewski2015catching,keefer2020visualizing,keefer2021selective} was proposed to detect the electronic coherence generated by the CI. In past decades, high-harmonic spectroscopy (HHS) has matured into a powerful approach to study the structure and dynamics of molecules \cite{kanai2005quantum,Mairesse2010,vozzi2011generalized,Frumker2012,Wong2013,Baykusheva2016,Marcelo2017PhysRevA,Uzan2020,peng2019attosecond}. For example, H. J. W{\"o}rner {\textit{et al.}} reported the application of HHS to detect CI dynamics \cite{worner2011conical,kraus2012time}.

While both CI and AC can induce a nonradiative transition, most of the previous works did not provide a specific recipe for discriminating between the CI and the AC. The main topological feature of CI which distinguishes it from the AC is that the wavepacket accumulates the geometric phase (GP) as it propagates around the CI. As early as 1963, Herzberg and Longuet-Higgins \cite{herzberg1963intersection} showed that the electronic wave function changes its sign for any closed path in the nuclear parameter space which encircles a CI. Later, in 1984, Berry pointed that this sign change is a special case of a more general GP factor \cite{berry1984quantal}, often referred as a ``Berry Phase". Further applications of GP include the hydrogen exchange reaction \cite{juanes2005theoretical,yuan2018observation} and the dissociation spectrum \cite{abe2006geometric,nix2008observation,bouakline2014investigation,xie2018signatures}. Since the high-harmonic generation (HHG) is driven by the ultrashort laser fields, and the HHS is particularly sensitive to the variation of ionization potential along with the population and the coherence between the superposed states, the HHS is an ideal tool to simultaneously monitor the ultrafast dynamics around the CI or the AC and to identify GP effect.
\begin{figure*}[htbp]
	\includegraphics[width=7in,angle=0]{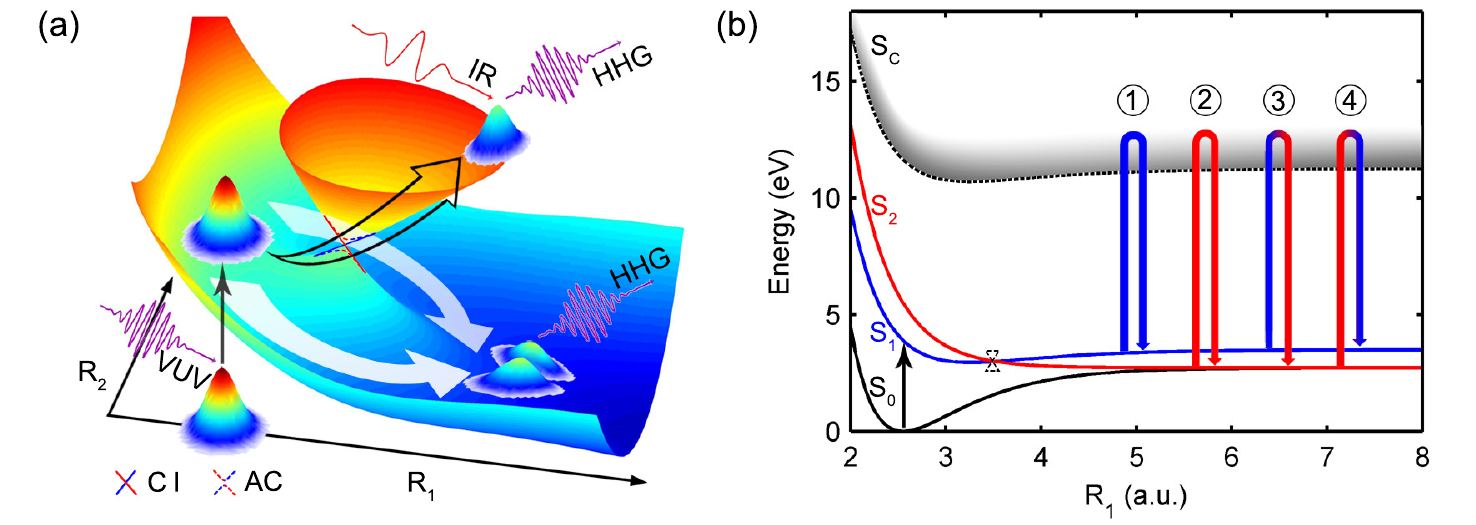} %3.5in
	\centering
	\caption[width=3in,angle=0]{(a) The adiabatic PESs and a schematic diagram of the pump-probe process. The wavepacket is pumped to excited state $ S_1 $ by the VUV pulse. Then, part of it passes through the CI (AC) (shown by the black hollow arrow), and the rest is divided into two parts bypassing the CI (AC) in the clockwise and the counterclockwise directions (shown by white translucent arrow). Finally, the HHG of the excited molecule is obtained with time-delayed IR pulse. (b) 1D slices of the diabatic PESs at $ R_2 = 0 $. The dashed funnel indicates the location of the CI (AC). The arrows marked by \ding{172}, \ding{173}, \ding{174}, \ding{175} corresponds to the four different channels in the HHG process.}\label{Fig_1}
\end{figure*} 

\begin{figure*}[tp]
	\includegraphics[width=7in,angle=0]{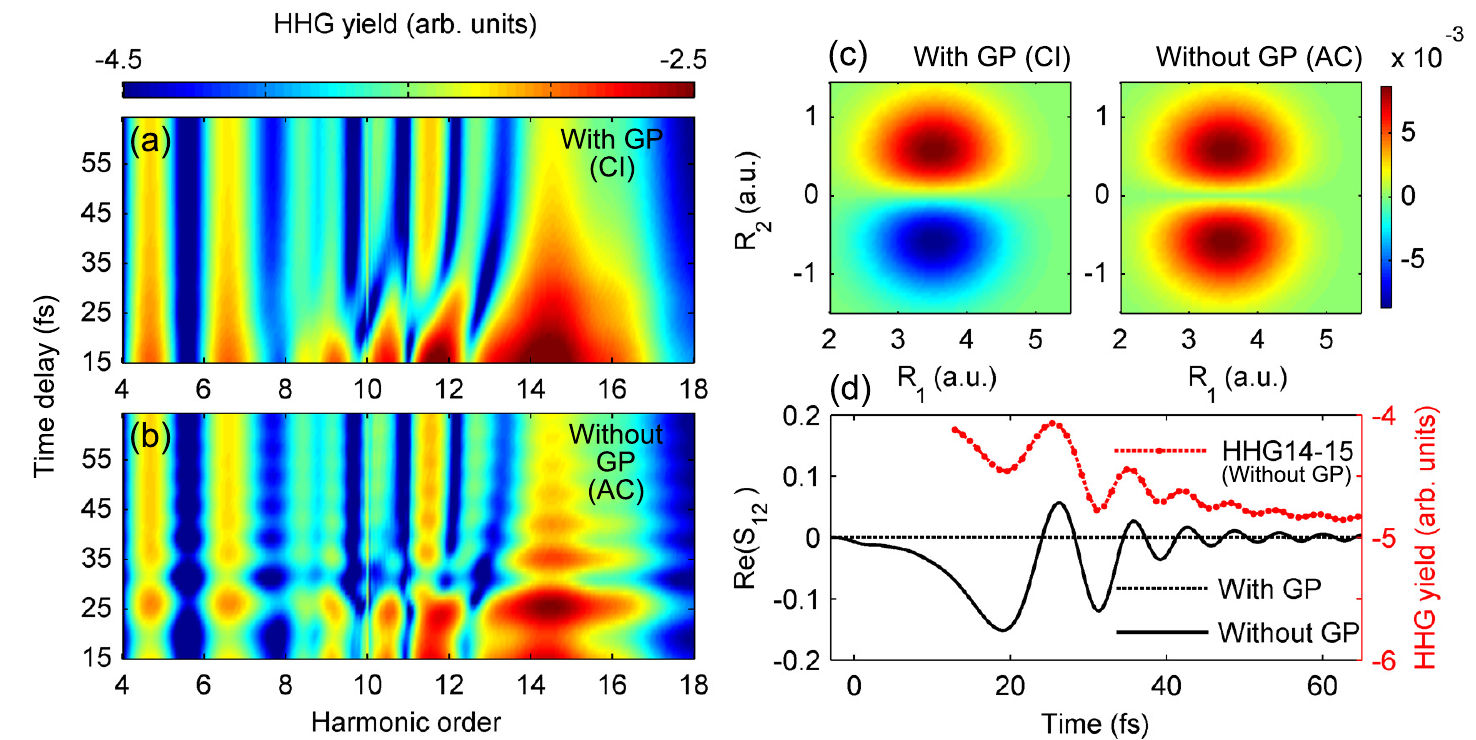} %3.5in
	\caption{The HHS after taking logarithm as a function of time delay (a) with and (b) without GP. The peak of the VUV pulse is set as the time origin. Time delay corresponds to the interval between the peak of the IR and the origin. Harmonic order is taken relative to the IR single photon energy. 
	(c) 2D plots for the NAC element $ V_{12} $ (with GP) and $ V^{\prime}_{12} $ (without GP). 
	(d) Real part of time dependent overlap integral $ S_{12} $ with (black dotted line) and without (black solid line) GP in the absence of IR laser. The intensity of delay-dependent harmonics (red dotted line) is obtained by integrating harmonics between the 14-15th order range in Fig. 2(b).}\label{Fig_2}
\end{figure*}

In this letter, we apply pump-probe HHS driven by a few-cycle IR pulse to distinguish the CI from the AC. Fig. \ref{Fig_1}(a) shows a schematic of the pump-probe process where the wavepacket initially located on the ground state $S_0$ is excited to state $S_1$ by an ultrafast VUV pulse. The nuclear wavepacket will dissociate along two branches encircling the CI or AC as indicated by the white arrows in Fig. \ref{Fig_1}(a). Meanwhile, part of the wavepacket passes through the CI or AC. Finally, a delayed strong IR probe laser is sent to generate HHS. To model the pump-probe HHG dynamics, the time-dependent Schrödinger equation (TDSE) in diabatic representation is solved (Atomic units are used unless otherwise indicated):
\begin{equation}\label{Eq_1}
	i\frac{\partial }{{\partial t}}
	\begin{pmatrix}
		\chi _0 (\mathbf R,t) \\ \chi _1 (\mathbf R,t) \\ \chi _2 (\mathbf R,t) \\ \chi _C (\mathbf R,\mathbf k,t)
	\end{pmatrix}
	= \left( {\hat{T} + \hat{H}_{el}} \right) 
	\begin{pmatrix}
		\chi _0 (\mathbf R,t) \\ \chi _1 (\mathbf R,t) \\ \chi _2 (\mathbf R,t) \\ \chi _C (\mathbf R,\mathbf k,t)
	\end{pmatrix},
\end{equation}
with Hamiltonian $ \hat{H}_{el} $ defined as follow,
\begin{equation}\label{Eq_2}
	\begin{bmatrix}
		E_{0}(\mathbf R) & d_{01} \mathbf E(t) & 0 & 0 \\ d_{01}^{*} \mathbf E(t) & E_{1}(\mathbf R) & V_{12}(\mathbf R) & D_{1 C}(\mathbf R, \mathbf k, t) \\ 0 & V_{12}^{*}(\mathbf R) & E_{2}(\mathbf R) & D_{2 C}(\mathbf R, \mathbf k, t) \\ 0 & D_{1 C}^{*}(\mathbf R, \mathbf k, t) & D_{2 C}^{*}(\mathbf R, \mathbf k, t) & \tilde{E}_{C}(\mathbf R, \mathbf k, t)
	\end{bmatrix}.
\end{equation}
Here, $ \hat{T} $ is the nuclear kinetic energy operator, $ \mathbf E(t) $ is the electric field. $ \chi _n $ is the nuclear wave function moving on state $ S_{n} $. $ E_{n} $ and $ E_{C} $ represent the potential energy surfaces (PESs) of neutral bound states $ S_n $ and ionic ground state $ S_C $, respectively. In the following simulation, the $E_{C}$ is assumed to be the same as $E_1$, except a vertical energy difference. $ \tilde E_C (\mathbf R, \mathbf k, t) = E_C (\mathbf R) + (\mathbf k + \mathbf A(t))^2/2 $ is the equivalent PESs of continuum state $ S_C $, with the vector potential $\mathbf A(t)$ and the electronic momentum $\mathbf k$. $ D_{nC} (\mathbf R, \mathbf k, t) = \mathbf E(t) d_{nC} (\mathbf R,\mathbf k+ \mathbf A(t)) $ is the product of the laser field $\mathbf E(t)$ and the transition dipole moments (TDMs) $ d_{nC} $ between the states $ S_n $ and $ S_C $. $ d_{nC} $ at each $ \mathbf R$ takes the form of the TDMs for hydrogen-like atoms \cite{lewenstein1994theory}. $ d_{01} $ is a constant TDM between states $S_0$ and $S_1$. $V_{12}$ is the nonadiabatic coupling (NAC) term between $S_1$ and $S_2$, and is related to the mixing angle $ \theta = 0.5arctan(2V_{12}/(E_{1}-E_{2})) $ \cite{xie2017nonadiabatic}. In the CI case, $V_{12}$ is anti-symmetric with respect to a mirror plane ($R_2=0$). Encircling the CI point with a closed loop $C$, $\theta$ evolves from $\pi/2$ to $-\pi/2$ \cite{ryabinkin2017geometric}. Since the Berry phase $ \gamma = \oint_C {\nabla \theta \cdot {\bf{ds}}} $ \cite{Baer1997,Baer2006ch3}, this phase shift of $\pi$ is considered as accumulation of GP. For the system with AC, the NAC term $V^{\prime}_{12}=|{V_{12}}|+\xi$ ($\xi$ is an arbitrarily small constant) is symmetric with respect to $R_2=0$. Assuming the same encirclement as before, $\theta$ changes from $\pi/2$ to $0$ and back to $\pi/2$, resulting in zero GP accumulation. Therefore the GP effect has been included with the anti-symmetric $V_{12}$, and has been excluded if we set $V^{\prime}_{12}=|{V_{12}}|$ \cite{Schuppel2020waveform}, as shown in Fig. \ref{Fig_2}(c).

The polarization can be recast as
\begin{equation}\label{Eq_3}
	\begin{aligned}
		P(t) &= \iint d \mathbf R d \mathbf k \quad[{\chi _1 (\mathbf R,t)\chi _C ^* (\mathbf R,\mathbf k,t)d_{1C} (\mathbf R, \mathbf k + \mathbf A(t)) }\\
		&+ {\chi _2 (\mathbf R,t)\chi _C ^* (\mathbf R,\mathbf k,t)d_{2C} (\mathbf R,\mathbf k + \mathbf A(t))}] + c.c.\,
	\end{aligned}
\end{equation}
after propagating the TDSE by the split operator algorithm based on fast Fourier transform. Finally, the HHS can be obtained by the Fourtier transformation of P(t)
\begin{equation}\label{Eq_4}
	I(\omega) \propto \omega ^4 \left| {\int_{ - \infty }^\infty {P(t)e^{i\omega t} dt} } \right|^2 .
\end{equation}

The parameters of the diabatic molecular model based on multiple 2-dimenional PESs are detailed in the Supplementary Material \cite{sm}. We employ the VUV pulse with the laser parameters of wavelength $ \lambda_1 = 318$ nm, peak intensity $ I_1 = 10^{13}$ W/cm$^2 $, and full width at half maximum (FWHM) $ \tau_1 = 3.18$ fs , which can pump about $ 70 \% $ of the electron population from the ground state $ S_0 $ to the excited state $ S_1 $. Regarding the delayed IR pulse, the parameters are wavelength $\lambda_2 = 1600$ nm, intensity $ I_2 = 10^{13}$ W/cm$^2 $, and FWHM $ \tau_2 = 5.33$ fs. The initial momentum of the ionized electron is set to $k_0 = 0$ instead of integrating over the entire range of momentum. The contribution of the ground state to the HHG has been neglected. The delay-dependent HHS were shown in Fig. \ref{Fig_2}(a) (includes the GP) and \ref{Fig_2}(b) (excludes the GP), in which the latter shows strong oscillations, while the former does not.

As the wavepacket passes through the CI or AC, electronic coherence can be built up and the system is excited with both population $\rho_{S_1}$, $\rho_{S_2}$, and coherences $\rho_{S_1S_2}$, $\rho_{S_2S_1}$ before the IR arrives. Due to the slow motion relative to the ultrafast IR pulse, the nucleus can be approximately regarded as fixed in the HHG process, namely the frozen-nuclear approximation (FNA). In the FNA and the strong-field approximation (SFA) \cite{lewenstein1994theory}, the electronic polarization for the fixed nuclear coordinate $ R_0 $ and time delay $ \tau $ can be written as:
\begin{equation}\label{Eq_5}
	P^{\prime}(t,\tau;R_0)=\sum\limits_{m,n=1}^2 P^{\prime}_{mn}(t,\tau;R_0),
\end{equation}
where $P^{\prime}_{mn}$ are the contributions from different channels:
\begin{equation}\label{Eq_6}
	\begin{aligned}
		P^{\prime}_{mn}&(t,\tau;R_0)= \int d \mathbf k \int_\tau^t dt^{\prime}i \mathbf E(t^{\prime}) \alpha_{mn}(\tau;R_0) \\
		&\times d^*_{mC} (\mathbf k+\mathbf A(t);R_0) d_{nC}(\mathbf k + \mathbf A(t');R_0) \\ 
		&\times e^{ - i \int_{t^{\prime}}^t ({{( \mathbf k + \mathbf A(t^{\prime \prime}) )^2 }/{2} + {E_C ( R_0 ) - E_n ( R_0 )) dt^{\prime \prime}} }} + c.c., \\
	\end{aligned}
\end{equation}
with
\begin{equation}\label{Eq_7}
		\alpha_{mn}(\tau;R_0)=\chi_m^* ({\tau;R_0}) \chi_n ({\tau;R_0}) e^{i(\omega_{m}-\omega_{n})(t-\tau)}.
\end{equation}
$\alpha_{mn}$ is not dependent on ionization time $t^{\prime}$ and recombination time $t$ since the depletion of the excited states can be neglected \cite{lewenstein1994theory}. Eq. \eqref{Eq_6} is similar to the traditional SFA model except that it is weighted by the overlap $\alpha_{mn}$.
As indicated in Fig. \ref{Fig_1}(b), $P^{\prime}_{11}, P^{\prime}_{22}, P^{\prime}_{12}, P^{\prime}_{21}$ corresponds to channel \ding{172}, \ding{173}, \ding{174}, \ding{175} respectively. We can classify the HHG process into direct (\ding{172}, \ding{173}) and indirect (\ding{174}, \ding{175}) ionization-recombination channels. The initial and final states of direct (indirect) channel where the electron is ionized and recombined are the same (different). Ignoring the phase term and integrating Eq. \eqref{Eq_7} over $\mathbf R$, the overlap integral $S_{mn}=\int d \mathbf R \chi_m^* ({t; \mathbf R}) \chi_n ({t; \mathbf R})$ is obtained. Previously, many works used the overlap integral $S_{mn}$ to probe the nuclear dynamics with the sub-cycle resolution \cite{baker2006probing,le2012theory,kowalewski2015catching}. According to Eq. \eqref{Eq_7}, unlike the direct channels related to the population $\rho_{S_1}$ and $\rho_{S_2}$, the indirect ones, e.g. $n\neq m$, yield an off-diagonal coherence which results in the oscillating behavior with respect to the interpulse delay. 
\begin{figure}[tp]%[H][htbp]
	\includegraphics[width=3.5in,angle=0]{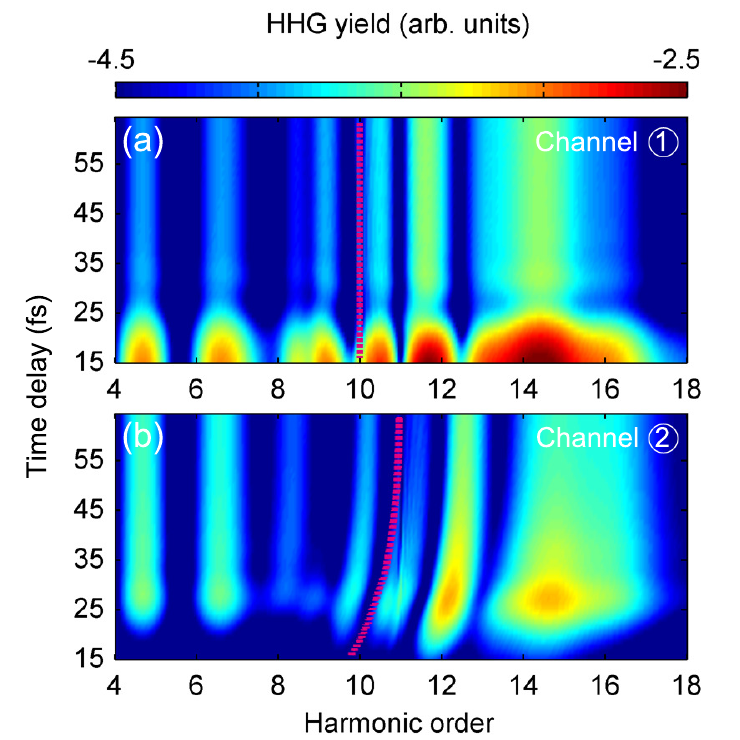} %3.5in
	\caption{Time-resolved HHS generated by the (a) channel \ding{172} and (b) channel \ding{173} with GP. The two time-varying energy difference $ E_C - E_1 $ and $ E_C - E_2 $ is plotted in (a) and (b) depicted by the dashed purple line. These are calculated at the coordinates of the two excited state wavepacket maximum positions, respectively.}\label{Fig_3}
\end{figure}

The next logical point that needs to be addressed is the eroding of oscillations in the CI case. The symmetric wavepackets in state $S_1$ will be changed to the anti-symmetric in state $S_2$ when it is transferred by the anti-symmetric NAC term $V_{12}$. Once the integral in the polarization (Eq. \eqref{Eq_6}) is calculated over the whole nuclear space, the contribution from the indirect channel will vanish. That is the reason that the oscillation feature disappear in HHS for the case of CI. This conclusion can be proved by Fig. \ref{Fig_2}(d), where the real part of $S_{12}$ with and without GP are ploted together with the time delay-dependent HHS intensity.

Another important phenomenon is that in both Fig. \ref{Fig_2}(a) and \ref{Fig_2}(b) the delay-dependent peaks of the spectrum split. Obviously, this phenomenon is not related to the GP effect, and originates rather from the direct channels because the indirect channels vanish in the CI case as explained above. Thus, the total harmonics in Fig. \ref{Fig_2}(a) can be well divided into channel \ding{172} and \ding{173}, by choosing a particular term in square brackets of Eq. \eqref{Eq_3}, as shown in Fig. \ref{Fig_3}. As the result, the time-resolved HHS from channel \ding{173} shows the blueshift while the contribution from the channel \ding{172} does not. According to Eq. \eqref{Eq_6}, the total phase of $q$-th order harmonic generated from the channel \ding{172} or channel \ding{173} can be written as \cite{chang1998temporal,Shin1999generation} 
\begin{align}\label{Eq_8}
	&\phi_q\left(t\right)=q\omega_{IR} t\nonumber\\
	&+\int_{t^{\prime}}^t\left({\left(\mathbf k+ \mathbf A\left(t^{\prime \prime}\right)\right)^2}/{2}+E_C\left(R_0\right)-E_n\left(R_0\right)\right) d t^{\prime \prime},
\end{align}
with $\omega_{IR}$ as the central frequency of IR laser.
Therefore, the instantaneous frequency of each harmonic is given by
\begin{align}\label{Eq_9}
	&\omega_q(t)=\frac{\partial \phi_q(t)}{\partial t}\nonumber\\
	&=q \omega_{IR} +{(\mathbf k+ \mathbf A(t))^2}/{2} +\left(E_C\left(R_0\right)-E_n\left(R_0\right)\right).
\end{align}
Here, the first and the second terms in the right hand side (RHS) depend solely on the driving laser field, which does not depend on the interpulse time delay. Thus, the blueshift originates from the third term in the RHS of Eq. \eqref{Eq_9}, e.g. the energy difference between the ionic ground state $S_C$ and the state $S_n$ from where the electron ionizes. In Fig. \ref{Fig_3}(a) and \ref{Fig_3}(b), the dashed purple lines are the delay-dependent energy differences for $E_C-E_1$ and $E_C-E_2$ at the nuclear coordinates corresponding to the maximum of the delay-dependent wavepackets. It is apparent that the energy differences are consistent with the blueshifts. The two paths result in the split of the total HHS in Fig. \ref{Fig_2}(a) and \ref{Fig_2}(b), which clearly indicates that the nuclear wavepacket passes through the CI or the AC at corresponding time delay of 20-30 fs.

Frequency modulation of HHS has been widely investigated previously, as summaried in Ref. \cite{bian2014probing}. Once the HHS originated from the leading (falling) edge of the laser envelop dominates, the final signal shows the blueshift (redshift). In our case, since the FNA is reasonable, the modulation of the ionization potential at the pulse duratoin timescale can be neglected. The frequency modulation contributed from the leading and falling edges can be neglected in the present work. In fact, the blueshift in the present case can be only observed in the case of few-cycle lasers. So, the time-resolution of the present blueshift is determined by the FWHM of the driving laser. 
\begin{figure}[tp]%[H]
	\includegraphics[width=3.5in,angle=0]{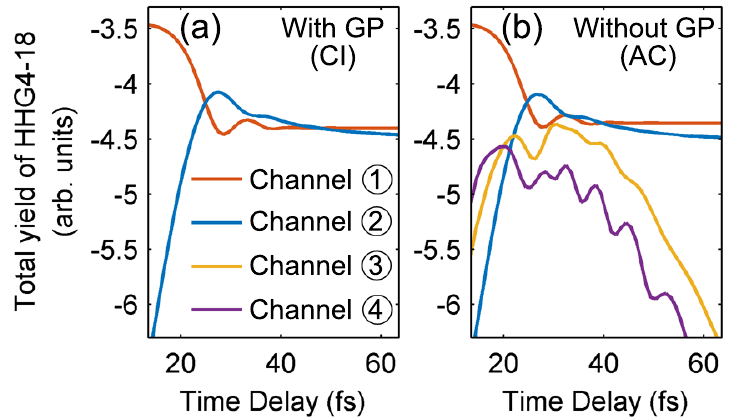} %3.5in
	\caption{Time-resolved HHS from the four channels (\ding{172}-\ding{175}) obtained after integrating Eq. \eqref{Eq_6} over $\mathbf R $ and integrating the harmonics in the 4th-18th order range. Panels (a) and (b) correspond to the results with and without GP effect, respectively. With the GP effect, the harmonic yields of channels \ding{174} and \ding{175} are around -19 orders of magnitude, and are thus omitted in panel (a).}\label{Fig_4}
\end{figure} 

Fig. \ref{Fig_4} depicts the time-resolved integrated intensity of HHS from the four channels, which is calculated using Eq. \eqref{Eq_6} and obtained by integrating over the space of $\mathbf R$. It indicates that the direct channels also show the weak oscillations, which are proportional to the populations of the excited states. These results are in full agreement with those obtained in Ref. \cite{worner2011conical}, in which the oscillation feature originating from the population dynamics has been observed by the transient grating technique. The contributions of the indirect channels are proportional to the coherence of the two excited states and thus show much stronger oscillations.

In summary, we show that the time-resolved HHS can probe directly the topological phase in CI and distinguish its quantitative behavior from the case of AC. When the nuclear wavepacket passes through the CI and AC, the electronic coherence may emerge as a consequence of NAC. The electronic coherence leads to the oscillatory feature in the HHS in the case of AC, while the vanishing oscillations in the case of CI are consequences of the symmetry breaking caused by the GP. Additionally, the HHS indicates the splitting of the signal into two peaks as the nuclear wavepackt passes through the CI or AC, which is attributed to the different frequency shifts of the HHS contributing by the different direct channels. The combine effects of peak splitting along the dynamical blueshift indicates a possibility to capture the real-time nonadiabatic molecular dynamics with the sub-femtosecond time resolution. Our result will benefit significantly from the ability to separately track electronic coherences and populations. A near future development of the multi-dimensional HHS \cite{jiang2020detecting,jiang2021multi} could be a potential tool to improve these results and track the dynamics of CI or AC with even higher resolution. 

This work was supported by National Natural Science Foundation of China (Grants No. 12074124, No. 11974185); Zijiang Endowed Young Scholar Fund, East China Normal University; Overseas Expertise Introduction Project for Discipline Innovation (B12024). S. J acknowledges the support by the start-up funding from East China Normal University. G. Y. is very grateful to Zhanjie Gao, Junjie Chen, Tong Wu, Lihan Chi, and Chen Qian for their help and discussions.

%\bibliographystyle{apsrev4-1}	
%\bibliography{bibfile}


\end{document}

\preprint{APS/123-QED}
	
	\title{Supplemental Material: Catching the geometric phase effect around conical intersection in molecules by high order harmonic spectroscopy}
	
	\author{Guanglu Yuan}
	\affiliation{Institute of Ultrafast Optical Physics, MIIT Key Laboratory of Semiconductor Microstructure and Quantum Sensing, Department of Applied Physics, Nanjing University of Science and Technology, Nanjing 210094, China}
	\affiliation{State Key Laboratory of Precision Spectroscopy, East China Normal University, Shanghai 200062, China}
	
	\author{Ruifeng Lu}
	\thanks{rflu@njust.edu.cn}
	\affiliation{Institute of Ultrafast Optical Physics, MIIT Key Laboratory of Semiconductor Microstructure and Quantum Sensing, Department of Applied Physics, Nanjing University of Science and Technology, Nanjing 210094, China}
	
	\author{Shicheng Jiang}
	\thanks{scjiang@lps.ecnu.edu.cn}
	\affiliation{State Key Laboratory of Precision Spectroscopy, East China Normal University, Shanghai 200062, China}
	
	\author{Konstantin Dorfman}
	\thanks{dorfmank@lps.ecnu.edu.cn}
	\affiliation{State Key Laboratory of Precision Spectroscopy, East China Normal University, Shanghai 200062, China}
	\affiliation{Collaborative Innovation Center of Extreme Optics, Shanxi University, Taiyuan, Shanxi 030006, China}
	\affiliation{Himalayan Institute for Advanced Study, Unit of Gopinath Seva Foundation, MIG 38, Avas Vikas, Rishikesh, Uttarakhand 249201, India}
	
	\date{\today}% It is always \today, today,
	% but any date may be explicitly specified
	
	%\begin{abstract}
		%In polyatomic molecules, 
	%\end{abstract}
	
	\maketitle
	%\tableofcontents
	This Supplemental Material contains the derivation of the time-dependent Schrödinger equation (TDSE) of a laser driven molecule and the parameters for the two-dimensional (2D) model based on multiple potential energy surfaces (PESs) in the diabatic representation. Furthermore, it presents the simulations for the delay-dependent high harmonic spectroscopy (HHS) of different channels calculated based on the frozen-nuclear approximation (FNA) and the strong-field approximation (SFA).
	
\section{the TDSE in diabatic representation}
	
	We use the Schrödinger equation to describe the properties and time-evolution of a molecular system driven by electric field $E\left( t \right)$ , 
	%The vector is not bold to avoid confusion with the matrix
	\begin{equation}\label{Eq_s1}
		i\frac{\partial }{{\partial t}}\Psi \left( {R,r,t} \right) = \left[ {\hat T \left( R \right) + \hat H_{e} \left( {r;R} \right) + r \cdot E\left( t \right) } \right]\Psi \left( {R,r,t} \right),
	\end{equation}
    where $\hat T = - \frac{1}{{2M}}\nabla ^2 $ is the nuclear kinetic energy operator and $ \hat H_{e} \left( {r;R} \right) = \hat T_e \left( r \right) + \hat U\left( {r;R} \right) $ is the clamped-nucleus Hamiltonian with the electronic kinetic energy operators $\hat T_e \left( r \right)$ and the potential energy function $\hat U\left( {r;R} \right)$. The eigenvalues $V$ and eigenfunctions $ \Phi $ of $ \hat H_{e} $ can be found from
    \begin{equation}\label{Eq_s2}
    	\hat H_{e} \left( {r;R} \right)\Phi _m \left( {r;R} \right) = V_m \left( R \right)\Phi _m \left( {r;R} \right).
    \end{equation}
    For a continuum state wave function $\Phi _k$ with electron momentum $k$
    \begin{equation}\label{Eq_s3}
    	\begin{aligned}
    	\hat H_{e} \left( {r;R} \right)\Phi _k \left( {r;R} \right) &= \left[ {\hat T_e \left( {r_{n - 1} } \right) + \hat U\left( {r_{n - 1} ;R} \right) + \hat T_e \left( {r'} \right) + \hat U\left( {r';R} \right)} \right]\Phi _k^{n - 1} \left( {r_{n - 1} ;R} \right)\Phi ^\prime _k \left( {r';R} \right) \\ 
    	&\approx \left( {V_k \left( R \right) + \frac{{k^2 }}{2}} \right)\Phi _k \left( {r;R} \right) = \tilde V_k \left( R \right)\Phi _k \left( {r;R} \right), \\ 
        \end{aligned}
    \end{equation}
    in which
    \begin{equation}\label{Eq_s4}
    \left[ {\hat T_e \left( {r'} \right) + \hat U\left( {r';R} \right)} \right]\Phi ^\prime _k \left( {r';R} \right) \approx \frac{{k^2 }}{2}\Phi ^\prime _k \left( {r';R} \right), \\
    \end{equation}
     as the role of $\hat U\left( {r';R} \right)$ is less pronounced for the ionized electron than that of strong laser field \cite{lewenstein1994theory}. In Eq. \eqref{Eq_s3}, it is approximately regarded that the ionized electron moves independently and has no electron correlation with the rest $n-1$ electrons.
     
     By using the eigenfunctions of $ \hat H_{e} $ as a basis set, the total wavefunction $\Psi$ can be expand as
     \begin{equation}\label{Eq_s5}
     \Psi \left( {R,r,t} \right) = \sum\limits_m {{\rm X}_m \left( {R,t} \right)\left| {\Phi _m \left( {r;R} \right)} \right\rangle } + \int {d k{\rm X}_k \left( {R,k,t} \right)\left| {\Phi _k \left( {r;R} \right)} \right\rangle } ,
     \end{equation}
     where $\rm X $ are nuclear functions that act as the expansion coefficients. Substitute Eq. \eqref{Eq_s5} into Eq. \eqref{Eq_s1}, and multiply $\left\langle {\Phi _l } \right|$ or $\left\langle {\Phi _{k'} } \right| $ from left on both sides to get
     \begin{equation}\label{Eq_s6}
     i\dot {\rm X}_l = \left[ {T + V_l } \right]{\rm X}_l - \frac{1}{{2M}}\sum\limits_m {\left( {2F_{lm} \nabla + G_{lm} } \right)} {\rm X}_m + E\left( t \right)\left( {\sum\limits_m {\mu _{lm} {\rm X}_m } + \int {dk\mu _{lk} {\rm X}_k } } \right),
     \end{equation}
     \begin{equation}\label{Eq_s7}
     	i\dot {\rm X}_{k'} = \left[ {T + \tilde V_{k'} } \right]{\rm X}_{k'} + E\left( t \right)\sum\limits_m {\mu _{mk'}^ * {\rm X}_m } + iE\left( t \right)\frac{\partial }{{\partial k}}{\rm X}_{k'}. 	
     \end{equation}
     In Eq. \eqref{Eq_s6} and Eq. \eqref{Eq_s7}, we define the transition dipole moment (TDM)
     $\mu _{lm}  = \left\langle {\Phi _l } \right|r\left| {\Phi _m } \right\rangle $, the derivative coupling $F_{lm}  = \left\langle {{\Phi _l }}
     \mathrel{\left | {\vphantom {{\Phi _l } {\nabla \Phi _m }}}
     	\right. \kern-\nulldelimiterspace}
     {{\nabla \Phi _m }} \right\rangle $, the scalar coupling $G_{lm}  = \left\langle {{\Phi _l }}
     \mathrel{\left | {\vphantom {{\Phi _l } {\nabla ^2 \Phi _m }}}
     	\right. \kern-\nulldelimiterspace}
     {{\nabla ^2 \Phi _m }} \right\rangle $, and $F_{lk} ,F_{k'k} ,G_{lk} ,G_{k'k} ,\mu _{k'k} $ are not included. Substitute $ K\left( t \right) = A\left( t \right) + k =  - \int {E\left( t \right)dt + k}$ for $k^\prime$ to rewrite Eq. \eqref{Eq_s7}, we can get
     \begin{equation}\label{Eq_s8}
     i\dot {\rm X}_K  = \left[ {T  + \tilde V_K } \right]{\rm X}_K  + E\left( t \right)\sum\limits_m {\mu _{mK}^ *  {\rm X}_m } .
     \end{equation}
     Using the relationship $G_{lm}  = \sum\limits_j {F_{lj} F_{jm} }  + \nabla F_{lm} $ (${\bf{G}} = {\bf{FF}} + \nabla {\bf{F}}$ in matrix notation) \cite{worth2004beyond}, it is possible to rewrite the nuclear Schrödinger equation, Eq. \eqref{Eq_s6} and Eq. \eqref{Eq_s8}, in matrix notation as
     \begin{equation}\label{Eq_s9}
     i\frac{\partial }{{\partial t}}{\boldsymbol{{\rm X}}} = \left[ { - \frac{1}{{2M}}\left( {\boldsymbol{\nabla}  + {\bf{F}}} \right)^2  + {\bf{V}} + {\boldsymbol{\upmu }}E\left( t \right)} \right]{\boldsymbol{{\rm X}}},
     \end{equation}
     with the dressed kinetic energy operator $\left( {\boldsymbol{\nabla}  + {\bf{F}}} \right)^2 $.
     
     The adiabatic picture, which has been used up to now, is transformed into the diabatic
     picture by a unitary transformation ${\bf{X} = \bf{A} \boldsymbol{\chi} }$ with diabatic wavefunctions $\boldsymbol{\chi}$, such that after multiplying by ${\bf{A}}^{\dag}$ the Eq. \eqref{Eq_s9} can be converted to
     \begin{equation}\label{Eq_s10}
        \begin{aligned}
     	i\frac{\partial }{{\partial t}}{\boldsymbol{\chi}} &= \left[ { - \frac{1}{{2M}} \boldsymbol{\nabla} ^2  + {\bf{A}}^{\dag} {\bf{VA}} + {\bf{A}}^{\dag} {\boldsymbol{\upmu }} {\bf{A}} E\left( t \right)} \right]{\boldsymbol{\chi}} \\ 
     	&= \left[ { - \frac{1}{{2M}} \boldsymbol{\nabla}^2  + {\bf{E}} + {\bf{d}}E\left( t \right)} \right]{\boldsymbol{\chi}}. \\
        \end{aligned}
    \end{equation}
    %&= \left[ { \hat{T}  + {\bf{E}} + {\bf{d}}E\left( t \right)} \right]{\boldsymbol{\chi}}. \\ 
    In Eq. \eqref{Eq_s10}, $\bf{A}$ is selected to satisfy $\nabla {\bf{A} + \bf{FA}} = 0$, so that $\left( {{\boldsymbol{\nabla}} + {\bf{F}}} \right)^2 {\bf{A}}{\boldsymbol{\chi}} = {\bf{A}} {\boldsymbol{\nabla}}^2 {\boldsymbol{\chi}}$ \cite{Baer2006ch2}. The adiabatic PESs ${\bf{V}}$ and TDMs ${\boldsymbol{\upmu }}$ are converted into the diabatic PESs ${\bf{E}}$ and TDMs ${\bf{d}}$. If only 3 bound states and continuum state are selected in Eq. \eqref{Eq_s5}, Eq. \eqref{Eq_s10} is reduced to Eq. 1 in the main text, where $\hat{H}_{el} = {\bf{E}} + {\bf{d}}E\left( t \right) $ is defined.

\section{MODEL SYSTEM}
    
    The 2D diabatic molecular model is completely specified by the Hamiltonian matrix $ \hat{H}_{el} $ and the nuclear kinetic energy operator $\hat{T}$ as a function of the internal coordinates $ R_1 $ and $ R_2 $, where the latter is a diagonal matrix with diagonal element given by
    \begin{equation}\label{Eq_s11}
    	\hat{T}_{nn}=-\frac{1}{2 M_{R_1}} \frac{\partial^2 }{{\partial^2 R_1 }}-\frac{1}{2 M_{R_2}} \frac{\partial^2 }{{\partial^2 R_2 }},   
    \end{equation}
    while the diabatic PESs in $ \hat{H}_{el} $ are given by
    \begin{equation}\label{Eq_s12}
    	\begin{aligned}
    		&E_{0}=A_1 e^{-a_1 R_1 + b_1}- B_1 e^{-c_1 R_1 + d_1 } + \frac{1}{2} K R_2^2 + e_1,\\
    		&E_{1}=A_2 e^{-a_2 R_1 + b_2}- B_2 e^{-c_2 R_1 + d_2 } + \frac{1}{2} K R_2^2 + e_2,\\
    		&E_{2}=A_3 e^{-a_3 R_1 + b_3} + \frac{1}{2} K R_2^2 + e_3,\\
    		&E_{C}=E_{1}+ e_4.\\
    	\end{aligned}    
    \end{equation}
    The TDM $ d_{01}=0.9 $. Although it is possible to use, for example, Dyson orbitals to calculate the bound-free TDMs $ d_{nC} $ for a more accurate description of the ionization process, choosing the form of hydrogen-like atoms does not affect our qualitative conclusion. For the system with conical intersection (CI), the nonadiabatic coupling term $V_{12}$ takes the form
    \begin{equation}\label{Eq_s13}
    	V_{12}=A_4 R_2 e^{-a_4 (R_1-b_4)^2} e^{-c_4 R_2^2 },
    \end{equation}
    while the nonadiabatic coupling term $V^{\prime}_{12}$ in the avoid crossing (AC) case takes the form
    \begin{equation}\label{Eq_s14}
    	V^{\prime}_{12}=A_4 (\lvert R_2 \rvert + \xi) e^{-a_4 (R_1-b_4)^2} e^{-c_4 R_2^2 }, 
    \end{equation}
    where the antisymmetric $V_{12}$ is converted to the symmetric $V^{\prime}_{12}$ by taking the absolute value of $ R_2 $ and adding an arbitrarily small constant $ \xi $. 
    As mentioned in the main text, the geometric phase (GP) effect has been included (excluded) in the $V_{12}$ ($V^{\prime}_{12}$).
    Here, $\xi$ is set to zero, which not only preserves the difference in GP, but also ensures that the adiabatic PESs is identical in both cases.
    The parameters of the model are as follows (in atomic units):
    \begin{equation}\label{Eq_s15}
    	\begin{aligned}
    		&M_{R_1}=30000,\ M_{R_2}=15000,\ A_1=0.1,\ A_2=0.02,\ A_3=1.0,\ A_4=0.025,\ B_1=0.2,\ B_2=0.04,\\
    		&K=0.04,\ a_1=3.0,\ a_2=2.4,\ a_3=2.4,\ a_4=1.2,\ b_1=7.65,\ b_2=7.8,\ b_3=3.84,\ b_4=3.5185, \\
    		&c_1=1.5,\ c_2=1.2,\ c_4=1.5,\ d_1=3.825,\ d_2=3.9,\ e_1=0.1,\ e_2=0.1285,\ e_3=0.1,\ e_4=0.285.\\
    	\end{aligned} 
    \end{equation}
     
     According to the above model, molecule is dissociating along the coordinate $ R_1 $ with CI standing in the way of dissociation pathway as indicated in Fig. 1 from the main text, which is found, for instance, in $ \rm{CH_2OO} $ \cite{samanta2014quantum}. 
     Similar molecular models have been used in other work, see Ref. \cite{Ferretti1996,cattaneo1997wave,Jadoun2021}.

\section{the delay-dependent HHS based on the FNA and the SFA}
\begin{figure}[htbp]%[H][htbp]
     	\includegraphics[width=7in,angle=0]{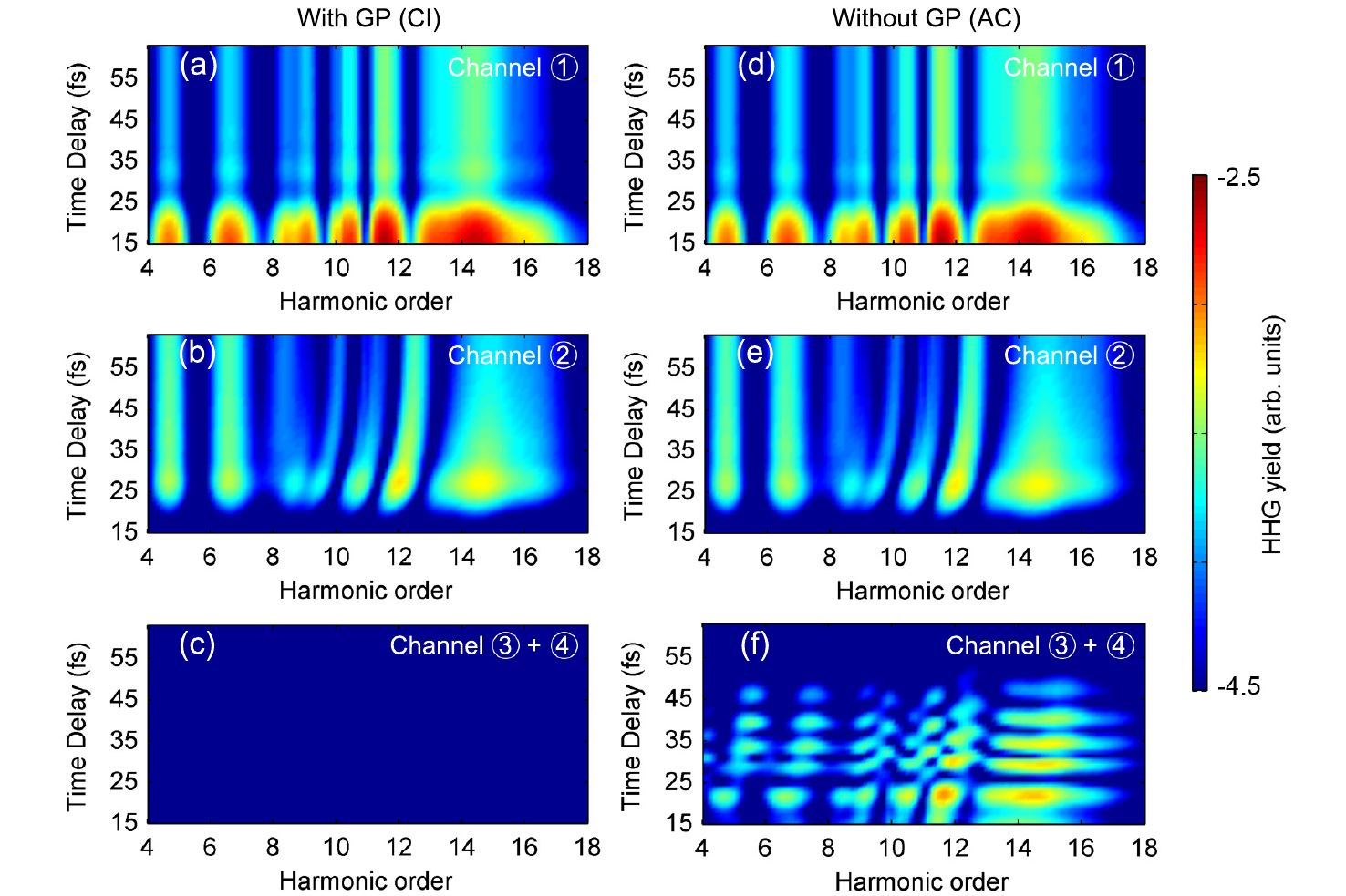} %3.5in
     	\caption{Time-resolved HHS from the four channels (\ding{172}-\ding{175}) obtained after integrating Eq. 6 from the main text over $\mathbf R $. Panels (a)-(c) and (d)-(f) correspond to the result with and without GP effect, respectively. With the GP effect, the harmonic yields of channels \ding{174} and \ding{175} are around -19 orders of magnitude, and are thus neglected in panel (c).}\label{Fig_s1}
     \end{figure}

     Fig. \ref{Fig_s1} depicts the time-resolved HHS from the four channels, which is calculated using Eq. 6 from the main text and obtained by integrating over the space of nuclear coordinates. Comparing the two cases with GP and without GP, the harmonics contributed by the direct channels (channels \ding{172} and \ding{173}) are approximately the same, and are also similar to the results calculated by TDSE in Figure 3 of the main text. In addition, the dynamical blueshift is also well reconstructed under the FNA and the SFA.
     
%\bibliographystyle{apsrev4-1}	
%\bibliography{bibfile}
